\documentclass[12pt]{article}

\def\QED{\mbox{\rule[0pt]{1.5ex}{1.5ex}}}
\usepackage[dvips]{graphics}

\begin{document}

\baselineskip 18pt

\title{\bf Digital Switching in the Quantum Domain}
\author{I.M. Tsai\footnote{E-mail : tsai@lion.ee.ntu.edu.tw}
\qquad and \qquad
        S.Y. Kuo\footnote{E-mail : sykuo@cc.ee.ntu.edu.tw}\\
        \\
        \em Department of Electrical Engineering,\\
        \em National Taiwan University,\\
        \em Taipei, Taiwan.\\
       }
\date{\small }
\maketitle

\begin{abstract}
In this paper, we present an architecture and implementation
algorithm such that digital data can be switched in the quantum
domain. First we define the connection digraph which can be used
to describe the behavior of a switch at a given time, then we show
how a connection digraph can be implemented using elementary
quantum gates. The proposed mechanism supports unicasting as well
as multicasting, and is strict-sense non-blocking. It can be
applied to perform either circuit switching or packet switching.
Compared with a traditional space or time domain switch, the
proposed switching mechanism is more scalable. Assuming an $n
\times n$ quantum switch, the space consumption grows linearly,
{\em i.e.} $O(n)$, while the time complexity is $O(1)$ for
unicasting, and $O(\log_2n)$ for multicasting. Based on these
advantages, a high throughput switching device can be built simply
by increasing the number of I/O ports.
\end{abstract}

\setlength{\parindent}{0.3 in}

\section{Introduction}

The demand for bandwidth is rapidly increasing due to the
explosive growth of network traffic. Networking technologies play
an important role in bridging the gap between limited resources
and the constantly increasing demand. In order to avoid a full
mesh architecture, a switching device is required to build a
realistic network. Over the past few years, a lot of enabling
technologies have emerged as candidates for achieving high
performance switching. Basically, switches act like automated
patch-panels, switching all the electrical or optical signals from
one port to another. Traditionally, digital switching can be done
in many ways. For example, by allocating physical separated paths,
switching can be done in the space domain. A 2-D MEMS optical
switch with precisely controlled micromirrors is essentially a
space domain switch. Similarly, by associating the data from each
port with a unique resource, switching can be performed in many
other ways, such as in the time domain, the wavelength domain, and
even a combination of these mechanisms.

On the other hand, quantum information science is a relatively new
field of study. Quantum computers were first discussed in the
early 1980's \cite{Ben80},\cite{Fey82},\cite{Deu85}. Since then, a
great deal of research has been focused on this topic. Remarkable
progress has been made due to the discovery of secure key
distribution \cite{Ben84}, polynomial time prime factorization
\cite{Sho94}, and fast database search algorithm \cite{Gro96}.
These results have recently made quantum information science the
most rapidly expanding research field. Other applications, such as
clock synchronization \cite{Joz00},\cite{Chu00}, and quantum
boolean circuit implementation \cite{Tsa01} have driven this field
further into the phase of real-world applications.

In this paper, we present a architecture and implementation
algorithm such that digital data can be switched in the quantum
domain. First we define the connection digraph which can be used
to describe the behavior of a switch at a given time, then we show
how a connection digraph can be implemented using elementary
quantum gates. The proposed mechanism supports unicasting as well
as multicasting and is strict-sense non-blocking \cite{Pat98}. It
can be applied to perform either circuit switching or packet
switching. Compared with a traditional space or time switch, the
proposed switching mechanism is more scalable. Assuming an $n
\times n$ quantum switch, the space consumption grows linearly,
{\em i.e.} $O(n)$, while the time complexity is $O(1)$ for
unicasting and $O(\log_2n)$ for multicasting. Based on these
advantages, a high throughput switching device can be built simply
by increasing the number of I/O ports.

\section{Notations and Preliminaries}

\subsection{Quantum State and Quantum Gates}

In a two-state quantum system, each bit can be represented using a
basis consisting of two eigenstates, denoted by $\vert 0 \rangle$
and $\vert 1 \rangle$ respectively. These states can be either
spin states of a particle ($\vert 0 \rangle$ for spin-up and
$\vert 1 \rangle$ for spin-down) or energy levels in an atom
($\vert 0 \rangle$ for ground state and $\vert 1 \rangle$ for
excited state). These two states can be used to simulate the
classical binary logic.

A classical binary logic value must be either {\bf \small ON} (1)
or {\bf \small OFF} (0), but not both at the same time. However, a
bit in a quantum system can be any linear combination of these two
states, so we have the state $\vert \psi
\rangle$ of a bit as
\begin{equation}
\vert \psi \rangle = c_0 \vert 0 \rangle + c_1 \vert 1
\rangle,\nonumber \label{superposition}
\end{equation}
where $c_0$, $c_1$ are complex numbers and $\vert c_0 \vert^2 + \vert c_1 \vert^2 = 1$.
In column matrices, this is written as
\begin{equation}
\vert \psi \rangle = \left( \begin{array}{c} c_0\\
c_1\end{array} \right)
\begin{array}{l}\\
.\end{array}
\end{equation}

The state shown above exhibits an unique phenomenon in quantum
mechanics called {\em superposition}. When a particle is in such a
superposed state, it has a part corresponding to $\vert 0 \rangle$
and a part corresponding to $\vert 1 \rangle$, at the same time.
When you measure the particle, the system is projected to one of
its basis ({\em i.e.} either $\vert 0 \rangle$ or $\vert 1
\rangle$). The overall probability for each state is given by the
absolute square of its amplitude. Taking the state $\vert \psi
\rangle$ in Eq.(\ref{superposition}) as an example, the
coefficient $\vert c_0 \vert^2$ and $\vert c_1 \vert^2$ represents
the probability of obtaining $\vert 0 \rangle$ and $\vert 1
\rangle$ respectively. Obviously, the sum of $\vert c_0 \vert^2$
and $\vert c_1 \vert^2$ will be $1$ to satisfy the probability
rule. To distinguish the above system from the classical binary
logic, a bit in a quantum system is referred to as a quantum bit,
or {\em qubit}.

Two or more qubits can also form a quantum system jointly. A
two-qubit system is spanned by the basis of the tensor product of
their own spaces. Hence, the joint state of qubit A and qubit B is
spanned by $\vert 00 \rangle_{AB}$, $\vert 01 \rangle_{AB}$,
$\vert 10 \rangle_{AB}$, and $\vert 11 \rangle_{AB}$, {\em i.e.}
\begin{equation}
\vert \phi \rangle_{AB} = c_0 \vert 00 \rangle_{AB} + c_1 \vert 01
\rangle_{AB} + c_2 \vert 10 \rangle_{AB} + c_3 \vert
11\rangle_{AB},
\end{equation}
where $c_0$, $c_1$, $c_2$, $c_3$ are all complex numbers and
$\vert c_0 \vert^2 + \vert c_1 \vert^2 + \vert c_2 \vert^0 + \vert
c_3 \vert^2 = 1$. In matrix form, this is equivalent to
\begin{equation}
\vert \phi \rangle_{AB} = \left( \begin{array}{c} c_0\\
c_1\\
c_2\\
c_3\end{array} \right)
\begin{array}{l}\\
\\
\\
.\end{array}
\end{equation}

The notations described above can be generalized to multiple-qubit
systems. For example, in a three-qubit system, the space is
spanned by a basis consisting of eight elements ($\vert 000
\rangle_{ABC}$, $\vert 001 \rangle_{ABC}$, \ldots , $\vert 111
\rangle_{ABC}$).

\begin{figure}[htbp]
 \center
 \scalebox{0.5}{\includegraphics{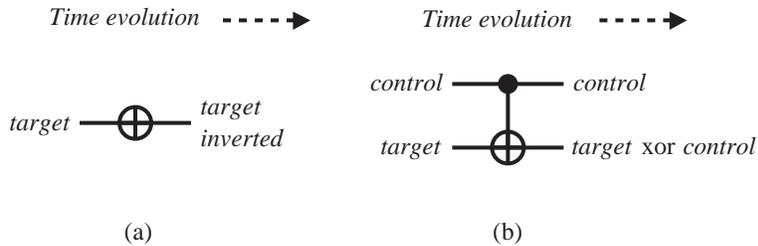}}
 \caption{The symbol and bit-wise operation for {\bf \small N} and {\bf \small CN} gate.}
 \label{figure1}
\end{figure}

A quantum system can be manipulated in many different ways, called
{\em quantum gates}. A quantum gate can be represented in the form
of a matrix operation. For example, a quantum {\em 'Not'} ({\bf
\small N}) gate applied on a single qubit can be represented by
multiplying a $2 \times 2$ matrix
\begin{equation}
N = \left( \begin{array}{c c} 0 & 1\\
1 & 0\end{array} \right)_, \label{not}
\end{equation}
which changes the state from $\vert 1 \rangle$ to $\vert 0
\rangle$ and from $\vert 0 \rangle$ to $\vert 1 \rangle$, as
\begin{equation}
N \cdot \left( \begin{array}{c}c_0\\
c_1\end{array} \right)=\left( \begin{array}{c c} 0 & 1\\
1 & 0\end{array} \right)\left( \begin{array}{c}c_0\\
c_1\end{array} \right)=\left( \begin{array}{c}c_1\\
c_0\end{array} \right)
\begin{array}{l}\\
.\end{array}
\end{equation}
The symbol of an {\bf \small N} gate is shown in
Fig.\ref{figure1}(a). Note that the horizontal line connecting the
input and the output is not a physical wire as in classical
circuits, it represents a qubit under time evolution.

Similarly, a two-bit gate can be represented by a $4 \times 4$
matrix. For example, a {\em 'Control-Not'} ({\bf \small CN}) gate
is represented by
\begin{equation}
CN = \left( \begin{array}{c c c c} 1 & 0 & 0 & 0\\
0 & 1 & 0 & 0\\
0 & 0 & 0 & 1\\
0 & 0 & 1 & 0\end{array} \right)
\begin{array}{l}\\
\\
\\
.\end{array} \label{control-not}
\end{equation}
The symbol of a {\bf \small CN} gate is shown in
Fig.\ref{figure1}(b). A {\bf \small CN} gate consists of one {\em
control} bit $x$, which does not change its value, and a {\em
target} bit $y$, which changes its value only if $x=1$. Assuming
the first bit is the control bit, the gate can be written as
$CN(\vert x,y \rangle)=\vert x,x \oplus y \rangle$, where
'$\oplus$' denotes exclusive-or. In matrix form, a {\bf \small CN}
gate changes the probability amplitudes of a quantum system as
follows:
\begin{equation}
CN \cdot \left( \begin{array}{c}c_0\\
c_1\\
c_2\\
c_3\end{array} \right)=
\left( \begin{array}{c c c c}1 & 0 & 0 & 0\\
0 & 1 & 0 & 0\\
0 & 0 & 0 & 1\\
0 & 0 & 1 & 0\end{array} \right)\left( \begin{array}{c}c_0\\
c_1\\
c_2\\
c_3\end{array} \right)=\left( \begin{array}{c}c_0\\
c_1\\
c_3\\
c_2\end{array} \right)
\begin{array}{l}\\
\\
\\
.\end{array}
\end{equation}

Further generalization of the quantum gates described above
involves {\em rotation} and {\em phase shift}. They control the
phase difference and relative contributions of the eigenstates to
the whole state. For example, a general single bit operation can
be represented using a matrix
\begin{equation}
U=\left(
\begin{array}{c c}e^{i(\delta+\frac{\alpha}{2}+\frac{\beta}{2})}\cos(\frac{\theta}{2}) &
e^{i(\delta+\frac{\alpha}{2}-\frac{\beta}{2})}\sin(\frac{\theta}{2})\\
-e^{i(\delta-\frac{\alpha}{2}+\frac{\beta}{2})}\cos(\frac{\theta}{2})
&
e^{i(\delta-\frac{\alpha}{2}-\frac{\beta}{2})}\sin(\frac{\theta}{2})\end{array}
\right)
\begin{array}{c}\\
.\end{array}
\label{onebit}
\end{equation}
This matrix can also be used to control the change between any two
probability amplitude components in a quantum system. Note that,
to satisfy the probability rule, all quantum gates $U$ in their
matrix form are unitary, {\em i.e.}
\begin{equation}
UU^\dagger = I,
\end{equation}
where $U^\dagger$ is the conjugate transpose of $U$.

Just like {\bf \small AND} and {\bf \small NOT} form a universal
set for classical boolean circuits, one- and two-bit gates are
sufficient to implement any unitary operation \cite{Div95},
\cite{Bar95.1}. A set of quantum gates which can be used to
implement any unitary operation is called a universal set. There
are many universal sets of one- and two-bit gates. A practical
approach is to use general one-bit rotation gates as in
Eq.(\ref{onebit}) and the {\bf \small CN} gate as a universal set.

\subsection{Qubit Permutation and Replication}

An important property regarding a quantum boolean operation is
that any quantum boolean logic can be represented using a {\em
permutation}. A permutation is a one-to-one and onto mapping from
a finite order set onto itself. A typical permutation $P$ is
represented using the symbol
\begin{eqnarray}
P = \left(
\begin{array}{cccccc}
a&b&c&d&e&f\\
d&e&c&a&f&b
\end{array}
\right)
\begin{array}{c}\\
.\end{array} \label{permutation}
\end{eqnarray}
This permutation changes $a$$\rightarrow$$d$, $d$$\rightarrow$$a$,
$b$$\rightarrow$$e$, $e$$\rightarrow$$f$, and $f$$\rightarrow$$b$,
with state $c$ remaining unchanged. A permutation can also be
expressed as disjoint {\em cycles}. A cycle is basically an
ordered list, which is represented as:
\begin{equation}
C=(e_1,e_2, \ldots ,e_{n-1},e_n). \label{cycle-rep}
\end{equation}
The order of the elements describes the operation. For example, in
Eq.(\ref{cycle-rep}), the cycle takes $e_1$$\rightarrow$$e_2$,
$e_2$$\rightarrow$$e_3$, \ldots ,$e_{n-1}$$\rightarrow$$e_n$, and
finally $e_n$$\rightarrow$$e_1$. The number of elements in a cycle
is called {\em length}. A cycle of length $1$ is called a {\em
trivial} cycle, which can be ignored as it does not change
anything. A cycle of length $2$ is called a {\em transposition}.
Using this notation, the same permutation $P$ shown in
Eq.(\ref{permutation}) can be written as
\begin{eqnarray}
P = (a,d)(c)(b,e,f) = (a,d)(b,e,f).
\end{eqnarray}

As we can see, a simple quantum boolean gate like {\bf \small CN}
can be regarded as a permutation, because the probability
amplitudes in the quantum state are manipulated in the same way.
In other words, a quantum boolean logic gate can be expressed as a
permutation, or cycles. For example, a {\bf \small CN} gate is
indicated by $P_{CN}=(10,11)$, changing $10$$\rightarrow$$11$ and
$11$$\rightarrow$$10$, leaving all other states unchanged.

\begin{figure}[htbp]
 \center
 \scalebox{0.5}{\includegraphics{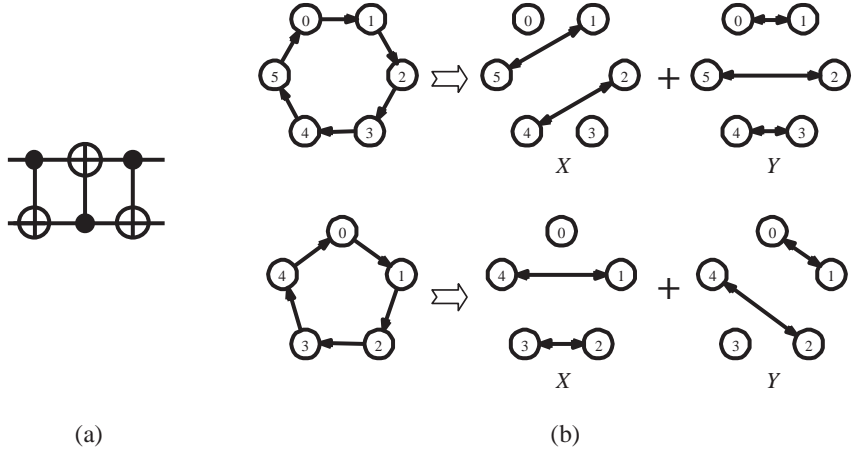}}
 \caption{The circuit for (a) a transposition and (b) general cycles.}
 \label{figure2}
\end{figure}

In addition to permute the probability amplitude of each
eigenstate, a qubit can be permuted as a whole. This is equivalent
to reshuffling the quantum states for each of the qubits. Since a
permutation can be decomposed into disjoint cycles, the
implementation actually consists of executing cycles of various
lengths in parallel. Because a cycle of length $1$ does not
permute anything, no circuit is required for a trivial cycle. For
a cycle of length $2$, the transposition can be done by three {\bf
\small CN} gates, as shown in Fig.\ref{figure2}(a).\\

The circuit is described as follows. For a two-qubit system
\begin{equation}
\vert \psi, \phi \rangle = c_{00} \vert 00 \rangle + c_{01} \vert
01 \rangle + c_{10} \vert 10 \rangle + c_{11} \vert 11 \rangle,
\end{equation}
the circuit transforms $\vert 00 \rangle \rightarrow \vert 00
\rangle$, $\vert 01 \rangle \rightarrow \vert 10 \rangle$, $\vert
10 \rangle \rightarrow \vert 01 \rangle$, and $\vert 11 \rangle
\rightarrow \vert 11 \rangle$. This is equivalent to the
permutation
\begin{equation}
P=(c_{00})(c_{01},c_{10})(c_{11}).
\end{equation}
Assuming the state of these two unentangled qubits are $\vert \psi
\rangle = \alpha \vert 0 \rangle + \beta \vert 1 \rangle$ and
$\vert \phi \rangle = \gamma \vert 0 \rangle + \delta \vert 1
\rangle$, where $\alpha,\beta,\gamma,\delta \in C$ and $\vert
\alpha \vert^2 + \vert \beta \vert^2 = \vert \gamma \vert^2 +
\vert \delta \vert^2 =1$, the joint state
\begin{equation}
\vert \psi \rangle \otimes \vert \phi \rangle = \alpha \gamma
\vert 00 \rangle + \alpha \delta \vert 01 \rangle + \beta \gamma
\vert 10 \rangle + \beta \delta \vert 11 \rangle
\end{equation}
is transformed to
\begin{eqnarray}
& & \alpha \gamma \vert 00 \rangle + \beta \gamma \vert 01 \rangle
+ \alpha \delta \vert 10 \rangle +
\beta \delta \vert 11 \rangle\\
& = & (\gamma \vert 0 \rangle + \delta \vert 1 \rangle) \otimes
(\alpha \vert 0
\rangle + \beta \vert 1 \rangle)\\
& = & \vert \phi \rangle \otimes \vert \psi \rangle,
\end{eqnarray}
which does the transposition. Note that once we have this basic
function, we can build a switching network in the same way as a
classical space switch. However, a more efficient implementation
exists, as will be presented later in this paper.

For a general $n$-qubit ($n \ge 3$) cycle $C=(q_0, q_1, q_2,
\cdots q_{n-1})$, it can be done by $6$ layers of {\bf \small CN}
gates without ancillary qubits \cite{Moo98}. The quantum
operations required to implement $C$ are shown below.

For an even $n$ ($n=2m$, $m=2,3 \ldots$), we define the following
non-overlapping qubit transpositions as:
\begin{eqnarray}
X & = & (q_{m-1},q_{m+1}) \cdots (q_2,q_{n-2})(q_1,q_{n-1}),\\
Y & = & (q_{m},q_{m+1}) \cdots (q_2,q_{n-1})(q_1,q_0).
\end{eqnarray}
The cycle can be implemented using
\begin{equation}
U=YX.
\end{equation}

On the other hand, for an odd $n$ ($n=2m+1$, $m=1,2,3 \ldots$), we
define the following non-overlapping qubit transpositions as:
\begin{eqnarray}
X & = & (q_{m},q_{m+1}) \cdots (q_2,q_{n-2})(q_1,q_{n-1}),\\
Y & = & (q_{m},q_{m+2}) \cdots (q_2,q_{n-1})(q_1,q_0).
\end{eqnarray}
Note that if the subscript $m+2 \ge n$ then $mod(m+2,n)$ is used
to avoid ambiguity. In the same way, the cycle can be implemented
using
\begin{equation}
U=YX.
\end{equation}
Two examples of $n=5$ and $n=6$ are shown in Fig.\ref{figure2}(b).

Note that both $X$ and $Y$ consist of disjoint transpositions and
can be executed in parallel using $3$ layers of {\bf \small CN}
gates, as shown in Fig.\ref{figure2}(a). As a result, each cycle
and the whole permutation can be performed using $6$ layers of
{\bf \small CN} gates. This achieves the constant time complexity
of a qubit permutation. If auxiliary qubits are used, a cycle can
be implemented using only $4$ layers of {\bf \small CN} gates
\cite{Moo98}.

\begin{figure}[htbp]
 \center
 \scalebox{0.5}{\includegraphics{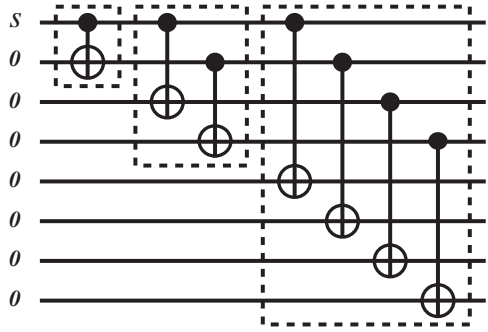}}
 \caption{An example of qubit replication from $s=\vert 0 \rangle$ or $\vert 1 \rangle$
to multiple targets.}
 \label{figure3}
\end{figure}

In addition to permutation, qubit replication ({\bf \small
FANOUT}) is also an important and non-trivial operation. Qubit
replication takes one bit as input and gives two copies of the
same bit value as output. In the classical world, we can do this
simply with a metallic contact, but it is well-known that quantum
mechanics does not allow us to make an exact copy of an unknown
qubit. This is called the quantum {\em non-cloning} theorem
\cite{Woo82}. However, if the source qubit is in either $\vert 0
\rangle$ or $\vert 1 \rangle$, the quantum state can be replicated
exactly using a {\bf \small CN} gate. For example, if $\vert \psi
\rangle$ is in either $\vert 0 \rangle$ or $\vert 1 \rangle$,
replicating $\vert \psi \rangle$ to the qubit $\vert \phi \rangle
= \vert 0 \rangle$ can be done simply by applying a {\bf \small
CN} gate with $\vert \psi \rangle$ as the control and $\vert \phi
\rangle$ as the target, {\em i.e.} $CN(\vert \psi, 0 \rangle)$.
Moreover, since both $\vert \psi \rangle$ and $\vert \phi \rangle$
can be used as the source qubits for further replication
processes, the number of copies will increase exponentially, which
allows $C$ copies of the same quantum state being replicated using
only $\log_2C$ layers of {\bf \small CN} gates, as shown in
Fig.\ref{figure3}. Note that the {\bf \small CN} gates which have
non-overlapping control and target qubits can be executed in
parallel and are grouped into one layer.

%

\section{Classical Digital Switching Techniques}

In classical digital communication, switching is needed in order
not to build a fully-meshed transmission network. Generally,
digital switching technologies fall under two broad categories:
{\em circuit switching} or {\em packet switching}. In this
section, we briefly introduce these two switching paradigms and
describe various implementations that can be employed to implement
the switching function. We also define the connection digraph
which can be used to illustrate the switching operation at a given
time.

\subsection{Digital Switching Networks}

\begin{figure}[htbp]
 \center
 \scalebox{0.48}{\includegraphics{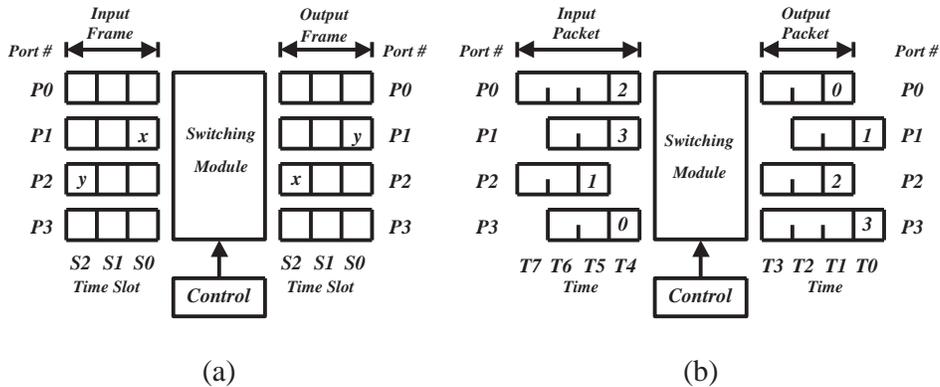}}
 \caption{Examples of (a) circuit switching and (b) packet switching.}
 \label{figure4}
\end{figure}

In circuit switching, a dedicated path or time slot is reserved
for an end-to-end bandwidth demand. The connection is established
at the time of call set-up and released when the call is torn
down. The function of the switching module is to transfer a
particular time slot in the input port to a time slot in the
output port. Assuming A (time slot $S0$ of port $P1$) and B (time
slot $S2$ of port $P2$) are making two-way communication via a $4
\times 4$ digital switch, as shown in Fig.\ref{figure4}(a). For
the connection from A to B, the switching module transfers the
data $x$ from $S0$ of $P1$ to $S2$ of $P2$. Similarly, for the
connection from B to A, it transfers the data $y$ from $S2$ of
$P2$ to $S0$ of $P1$. These operations complete the data exchange
between A and B.

Packet switching is more sophisticated than circuit switching.
Modern packet switching networks take packets that share the same
transmission line as input. A packet can have either a fixed or
variable length with a limited maximum size. When a packet arrives
at a node, it is stored first and then forwarded to the desired
node according to its header as shown in Fig.\ref{figure4}(b). For
example, assume each of the packets in Fig.\ref{figure4}(b) has
the destination port number as indicated in the header of the
packet. The switching module at time $T5$ needs to switch the data
from input port $P0$, $P1$, $P2$, and $P3$ to output port $P2$,
$P3$, $P1$, and $P0$ respectively.

Although significant differences such as {\em data dependency} and
{\em output contention} exist between circuit switching and packet
switching, they still have similarities. In both circuit switching
and packet switching, the control block needs to specify the
switching configuration for each individual time slot, so the data
in that particular time slot can be switched correctly. The
configuration describes how the I/O ports should be switched at a
given time. The actual switching operation depends on which
switching technique is used. There are many switching techniques
used today. Some of the basic switching techniques are described
in the following section.

\begin{figure}[htbp]
 \center
 \scalebox{0.5}{\includegraphics{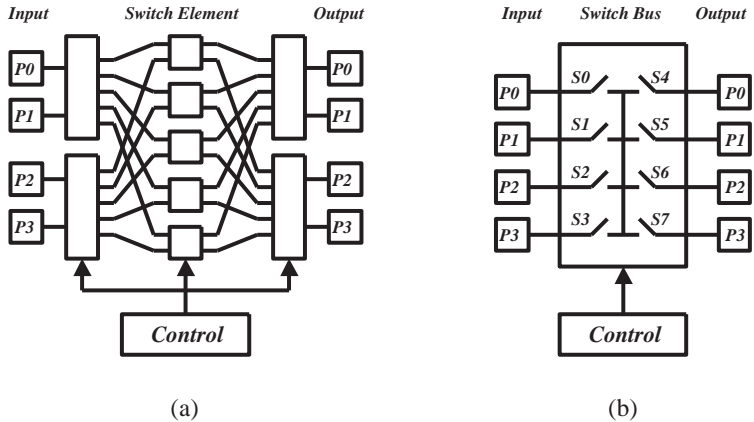}}
 \caption{Examples of classical digital switching techniques.}
 \label{figure5}
\end{figure}

In the field of classical digital switching, various techniques
have been used to switch the input data to the corresponding
output port. For example, data can be switched in the space
domain, the time domain, or the wavelength domain, etc. If the
data is switched in the space domain, {\em i.e.} space division
switching, usually a physical path or a dedicated time slot is
reserved to establish the connection. For example, in the crossbar
architecture, a rectangular array of cross-points serve as a
simple space switching architecture. Every output port can be
reached by every input port in a non-blocking way by closing a
single cross-point. A more sophisticated space division switch
utilizes multiple stages of rectangular arrays is shown in
Fig.\ref{figure5}(a).
A connection is established
by closing proper cross-points to select a path from the inlet to
the outlet \cite{Clo53}.

A device that switches the data in the time domain is called a
time division switch. Time division technology is widely used in
modern digital communication. In a time division switch,
connections are established in a time-sharing manner, so a
connection occupies the resources for only a short duration of
time. For example, in Fig.\ref{figure5}(b), the connection from
the inlet $P1$ to the outlet $P3$ is established by closing switch
$S1$ and $S7$. This process is executed for each of the
connections in a cyclic way to achieve switching functionality.
Primarily owing to the low cost of semiconductor devices, the
implementation of a time division switch is usually done by using
digital memory. Data received over an incoming port is written
into the memory, the switching is accomplished by reading out the
individual bits in the desired time slot, which is equivalent to
connecting the inlet to the outlet for data transfer.


\subsection{Connection Digraphs}

Before we describe how digital switching can be done in the
quantum domain, we define a {\em connection digraph} as follows:\\

{\bf Definition 1:} Given an $n \times n$ switch, the connection
digraph at time $t$, $G^t=\{V,E^t\}$, is a digraph such that
\begin{enumerate}
\item Each $v_i \in V (i=0,1,\ldots n-1)$ represents an I/O port.
\item $\overrightarrow{v_m v_n} \in E^t$ if and only if a connection exists from
the input port $v_m$ to the output port $v_n$ at time $t$.\hfill
\QED
\end{enumerate}

In a connection digraph, each node represents an I/O port, a
directed edge $\overrightarrow{v_m v_n}$ is used to describe a
connection when the connection from input port $v_m$ to output
port $v_n$ is active. The digraph describes the connection status
of the switch at a given time, and is called the connection
digraph at time $t$. Note that the directed edge
$\overrightarrow{v_m v_n}$ denotes only a one-way data path. For a
point-to-point two-way communication between $v_m$ and $v_n$, both
$\overrightarrow{v_m v_n}$ and $\overrightarrow{v_n v_m}$ have to
be used. Obviously, due to the connection set-up and torn-down
processes, the connection digraph is a function of time.

Depending on the status of the switch, the topology of a
connection digraph varies. In a general digraph, it is possible
that a node has multiple predecessors and multiple successors.
However, when there is no output contention or the problem is
solved elsewhere, each node will have at most one predecessor. As
to the number of successors, it depends on the type of the
connection. In a multicast connection, the source node has
multiple successors, while in a unicast connection, only a single
successor is possible. In the following sections, we will discuss
the connection digraph based on this model and show that any
connection digraph actually consists of a set of basic topologies
as disjoint sub-digraphs. These basic topologies are defined as
follows:\\

{\bf Definition 2:} Given a digraph $G=(V,E)$ with only one node,
{\em i.e.} $V=\{v\}$. $G$ is called a {\em null node} if
$E=\emptyset$. Otherwise $G$ is called a {\em loopback} when
$E=\{\overrightarrow{v v}\}$.
\hfill \QED\\

\begin{figure}[htbp]
 \center
 \scalebox{0.5}{\includegraphics{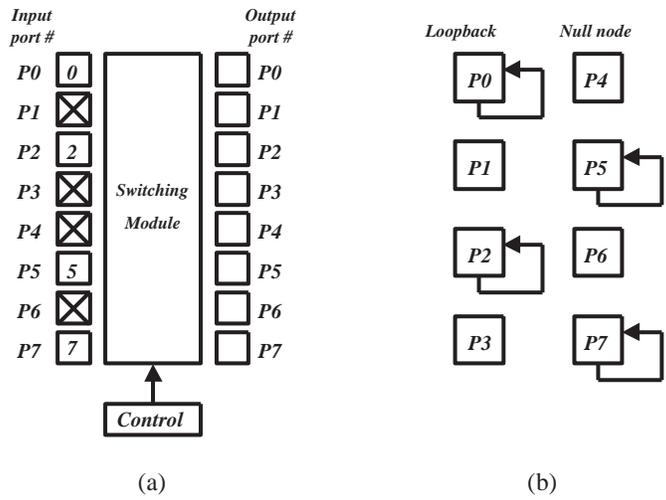}}
 \caption{A connection digraph with null nodes and loopbacks.}
 \label{figure6}
\end{figure}

In a connection digraph, a null node without predecessor and
successor means there is neither input traffic coming from that
port nor output traffic going to that port. For a port without
incoming traffic, we assume the stuff bits are all $0$'s. However,
a single node with a directed edge to itself means the input
traffic goes back to the same port. This trivial cycle effectively
denotes a loopback. A loopback $G^L$ can be made from a null node
$G^N$ simply by linking the null node to itself. $G^L$ is called
the extension loopback of $G^N$, denoted by $E(G^N)$. An example
consists of null nodes and loopbacks is shown in
Fig.\ref{figure6}(a). The numbers in the boxes represents the
destination port numbers. An 'X' represents no input traffic. Its
corresponding connection digraph is depicted in
Fig.\ref{figure6}(b).\\

{\bf Definition 3:} Given a connected digraph $G=(V,E)$ with $n$
($n \ge 2$) nodes. $G$ is called a {\em queue} if
\begin{enumerate}
\item there exists one and only one {\em head} $v_h \in V$, such that
for each $v_i \in V$, $\overrightarrow{v_i v_h} \notin E$.
\item there exists one and only one {\em tail} $v_t \in V$, such that
for each $v_i \in V$, $\overrightarrow{v_t v_i} \notin E$.
\item for each $v_i \in V (i \ne t)$, there exists one and only one
$v_j$, such that $\overrightarrow{v_i v_j} \in E$. \hfill \QED\\
\end{enumerate}

\begin{figure}[htbp]
 \center
 \scalebox{0.5}{\includegraphics{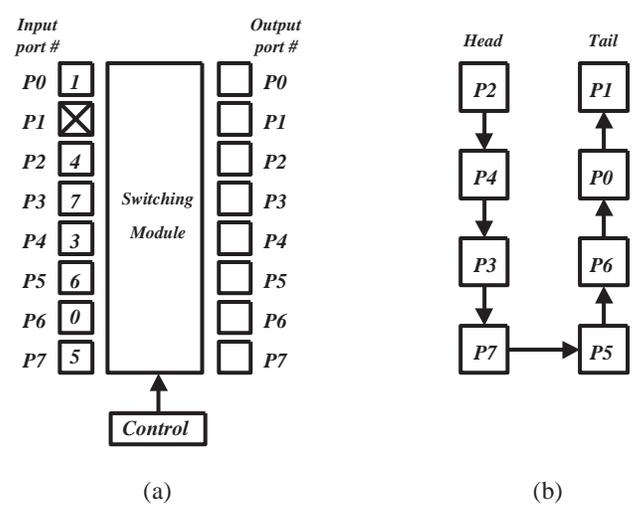}}
 \caption{An example of a queue connection and its connection digraph.}
 \label{figure7}
\end{figure}

A queue can be represented as a linear array from the head $v_h$
to the tail $v_t$, and is denoted as $[v_h, v_1, v_2, \ldots ,
v_{n-2}, v_t]$. This notation means the connection at a given time
includes $\overrightarrow{v_h v_1}$, $\overrightarrow{v_1 v_2}$,
\ldots, and $\overrightarrow{v_{n-2},v_t}$. Note that there is no
input traffic coming from $v_t$ and no output traffic going to
$v_h$. An example of a queue connection is shown in
Fig.\ref{figure7}(a), with its connection digraph
$G^Q=[P2,P4,P3,P7,P5,P6,P0,P1]$ depicted in Fig.\ref{figure7}(b).
Each connection in a queue is apparently a unicast connection,
because there is at most one outgoing arrow from each node.

Connecting the tail to the head of a queue forms a {\em cycle},
which is defined as follows:\\

{\bf Definition 4:} Given a connected digraph $G=(V,E)$ with $n$
($n \ge 2$) nodes, $G$ is called a cycle if
\begin{enumerate}
\item for each $v_i \in V$, there exists one and only one $v_j$,
such that $\overrightarrow{v_j v_i} \in E$.
\item  for each $v_i \in V$, there exists one and only one $v_k$,
such that $\overrightarrow{v_i v_k} \in E$. \hfill \QED \\
\end{enumerate}

\begin{figure}[htbp]
 \center
 \scalebox{0.5}{\includegraphics{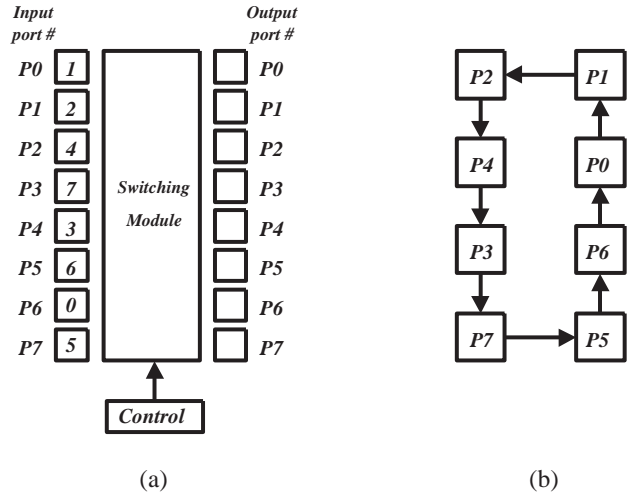}}
 \caption{An example of a cycle connection and its connection digraph.}
 \label{figure8}
\end{figure}

Using the same notation, a cycle connection is represented as
$(v_0, v_1, v_2, \ldots , v_{n-2}, v_{n-1})$. This means the
connection at a given time includes $\overrightarrow{v_0 v_1}$,
$\overrightarrow{v_1 v_2}$, \ldots ,$\overrightarrow{v_{n-2}
v_{n-1}}$, and $\overrightarrow{v_{n-1},v_0}$. In the case of a
cycle, each port has its input as well as output. As described
earlier, the tail and head of a queue $G^Q$ can be connected to
form a cycle $G^C$. $G^C$ is called the extension cycle of $G^Q$,
denoted by $E(G^Q)$ . An example of a cycle connection is shown in
Fig.\ref{figure8}(a), with its connection digraph
$G^C=(P2,P4,P3,P7,P5,P6,P0,P1)$ depicted in Fig.\ref{figure8}(b).

In order to describe a multicast connection, we define the
following connection digraphs:\\

{\bf Definition 5:} Given a connected digraph $G=(V,E)$ with $n$
($n \ge 2$) nodes, $G$ is called a {\em tree} if
\begin{enumerate}
\item there exists one and only one {\em root} $v_r \in V$, such that
for each $v_i \in V$, $\overrightarrow{v_i v_r} \notin E$.
\item there exists a collection of nodes $L$ called leaves, such that for each $v_l \in
L$ and $v_i \in V$, $\overrightarrow{v_l v_i} \notin E$.
\item for each $v_i \in V-L$, there exists at least one $v_j$, such
that $\overrightarrow{v_i v_j} \in E$. \hfill \QED\\
\end{enumerate}

\begin{figure}[htbp]
 \center
 \scalebox{0.5}{\includegraphics{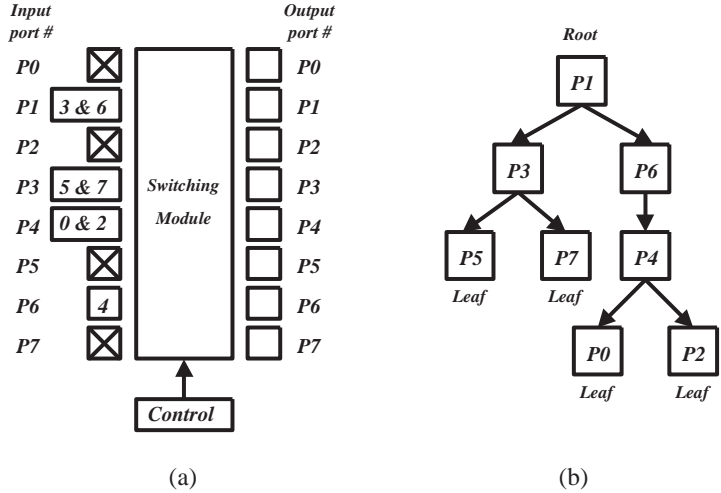}}
 \caption{An example of a tree connection and its connection digraph.}
 \label{figure9}
\end{figure}

The nodes in a tree can be divided into three categories: root,
internal nodes, and leaves. For the root, the output data is
directed to possibly multiple output ports, but no data goes to
the root. However, all leaves receive data without generating
traffic. All internal nodes have exactly one predecessor and at
least one successor. A tree can be represented as a concatenation
of queues like $G^T=[v^0_h, \ldots v^0_t][v^1_h, \ldots
v^1_t]\ldots[v^n_h, \ldots v^n_t]$, with $v^0_h$ be the root and
each of the $v^n_h$ ($n \ge 1$) be the tail of one of the previous
queues. An example of a tree connection is shown in
Fig.\ref{figure9}(a). If there are multiple numbers in a box, they
represent a multicast connection. Its corresponding connection
digraph $G^T=[P1][P1,P3][P1,P6,P4][P3,P5][P3,P7][P4,P0][P4,P2]$ is
depicted in Fig.\ref{figure9}(b). Note that a queue is a special
case of trees, with each node having only one successor.

Connecting any leaf to the root of a tree forms a {\em forest},
which is defined as follows:\\

{\bf Definition 6:} Given a connected digraph $G=(V,E)$ with $n$
nodes $(n \ge 2)$, $G$ is called a forest if
\begin{enumerate}
\item there is one and only one cycle $G^C=(V^C,E^C)$ exists as a
sub-digraph of $G$.
\item let $G^\prime=\{\overrightarrow{v_i v_j} \mid v_i \in V^C,
\overrightarrow{v_i v_j} \in E, \overrightarrow{v_i v_j} \notin
E^C$\}. $G-G^\prime$ contains the cycle $G^C$ and a collection of
disjointed null nodes, queues, and/or trees.
\item each $v_j$ is either one of the null nodes, the head of a queue,
or the root of a tree in $G-G^\prime$. \hfill \QED\\
\end{enumerate}

\begin{figure}[htbp]
 \center
 \scalebox{0.5}{\includegraphics{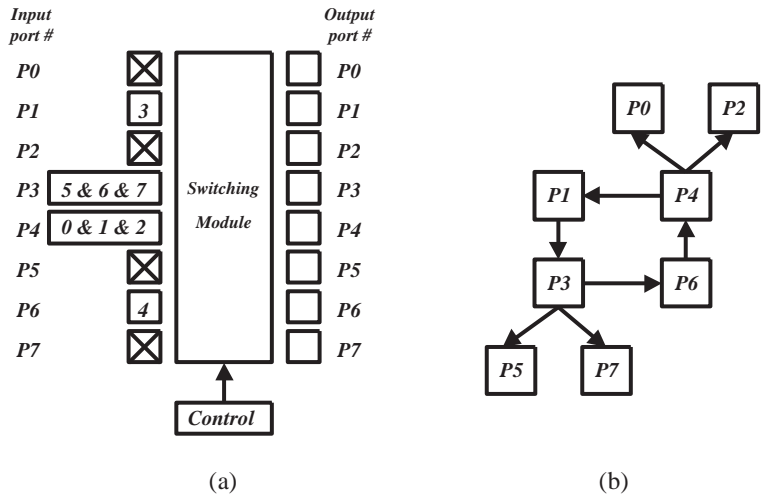}}
 \caption{An example of a forest connection and its connection digraph.}
 \label{figure10}
\end{figure}

A forest basically contains one and only one cycle $G^C=(V^C,E^C)$
as a sub-digraph, with some of its nodes linked to either a null
node, the head of a queue, or the root of a tree. Following this
structure, a forest can be represented by
$G^F=\{G^C,G^1,G^2,G^3\cdots\}$, where $G^1$, $G^2$, $G^3$, \ldots
be either a null node, a queue, or a tree. A forest can be
extended from a tree by connecting any leaf to the root. A forest
$G^F_l$ formed by connecting the leaf $l$ with the root of $G^T$
is called the extension forest of $G^T$, denoted by $E_l(G^T)$. An
example of a forest connection is shown in Fig.\ref{figure10}(a),
with its connection digraph
$G^F=\{(P4,P1,P3,P6),[P3][P3,P5][P3,P7],[P4][P4,P2][P4,P6]\}$
depicted in Fig.\ref{figure10}(b).

Since each node in a unicast connection has at most one successor,
a unicast connection digraph only consists of disjoint null nodes,
loopbacks, queues, and/or cycles as sub-digraphs. However, a
multicast connection switches the data from one node to multiple
successors, so a multicast connection digraph consists of disjoint
null nodes, loopbacks, queues, cycles, trees, and/or forests as
sub-digraphs. Based on these results, we describe the architecture
of quantum switching and show how it can be used to implement a
connection digraph in the next section.

\section{Switching in the Quantum Domain}

\subsection{Principle of Digital Quantum Switch}

\begin{figure}[htbp]
 \center
 \scalebox{0.5}{\includegraphics{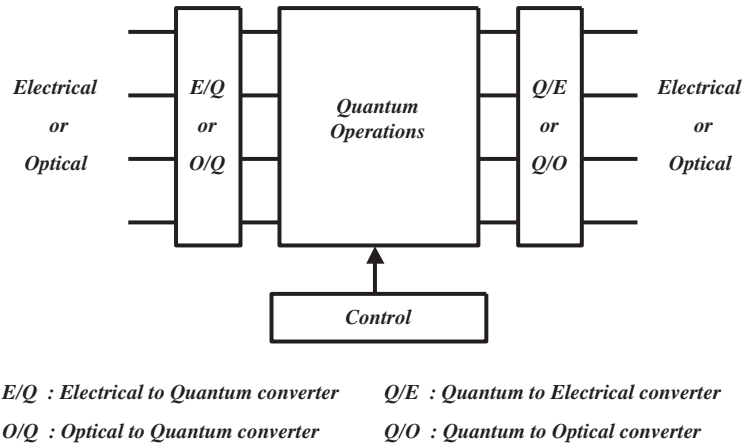}}
 \caption{The architecture of a digital quantum switch.}
 \label{figure11}
\end{figure}

The proposed architecture for building a digital {\em quantum
switch} is depicted in Fig.\ref{figure11}. To switch classical
digital data in the quantum domain, first we have to convert the
classical data into qubits. For example, in a quantum switch with
optical I/O ports, an optical to quantum converter (O/Q) is used
to convert optical input into qubits. In an O/Q, '$0$' is
converted into $\vert 0 \rangle$ and '$1$' is converted into
$\vert 1 \rangle$. This can be done by exciting an electron using
a light pulse of a certain frequency. All qubits are then permuted
({\em i.e.} switched) by the unitary operations under the
supervision of the control subsystem. After the permutation, all
qubits are converted back into their optical form by a quantum to
optical converter (Q/O). This can be done by measuring the qubits
to recover the original classical information.

\subsection{Connection Digraph Implementation}

In this section, we show how a connection digraph can be
implemented using {\bf \small CN} gates. First we describe the
connection digraph transformation guideline, then we demonstrate
how this guideline can be used to implement a connection digraph.
Both unicasting and multicasting will be covered in detail.

\subsubsection{Guideline for implementing a Connection Digraph}

\begin{figure}[htbp]
 \center
 \scalebox{0.4}{\includegraphics{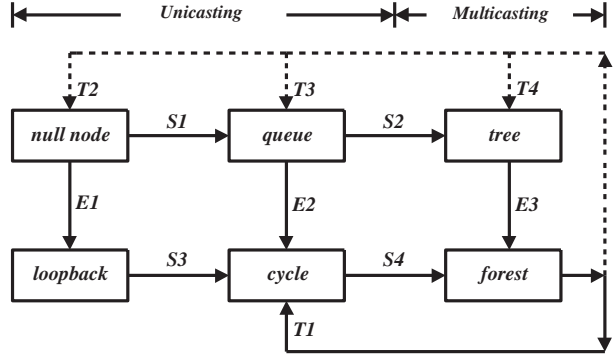}}
 \caption{Inter-related connection topologies.}
 \label{figure12}
\end{figure}

As described earlier, due to the nature of the connection,
unicasting and multicasting have different connection digraphs.
The digraph of a unicast connection has a collection of disjointed
null nodes, loopbacks, queues, and/or cycles as sub-digraphs.
However, in the digraph of a multicast connection, sub-digraphs
like trees and forests are possible. As a matter of fact, these
topologies are inter-related. This is shown in Fig.\ref{figure12}
and summarized as follows:
\begin{enumerate}
\item A null node can be regarded as a special case of a queue,
denoted by the arrow S1.
\item A queue can be regarded as a special case of a tree,
denoted by the arrow S2.
\item A loopback can be regarded as a special case of a cycle,
denoted by the arrow S3.
\item A cycle can be regarded as a special case of a forest,
denoted by the arrow S4.
\end{enumerate}

Of course, the binary relation "{\em is a special case of}" is
transitive, so a null node and a loopback are special cases of
tree and forest respectively. Fig.\ref{figure12} also shows the
binary relation "{\em can be extended to}" as follows:
\begin{enumerate}
\item A null node $G^N$ can be extended to a loopback $G^L=E(G^N)$, denoted by the
process E1.
\item A queue $G^Q$ can be extended to a cycle $G^C=E(G^Q)$, denoted by the
process E2.
\item A tree $G^T$ can be extended to a forest $G^F=E_l(G^T)$, denoted by the
process E3.
\end{enumerate}
Note that the process of extension only transfers the incoming
data from an idle inlet (all $0$'s) to an outlet which has no
outgoing traffic, this does not change the switching function.

The first step of our guideline for implementing a connection
digraph is to transform each disjointed sub-digraph into loopbacks
and/or cycles. Since no circuit is needed to implement a loopback
and only $6$ layers of {\bf \small CN} gates are sufficient to
implement a cycle, the switching can be done efficiently. Some of
these transformations are straightforward. For example, following
E1, a null node $G^N$ can be extended to a loopback $G^L=E(G^N)$.
Also, following E2, a queue $G^Q$ can be extended to a cycle
$G^C=E(G^Q)$. However, for a tree or a forest, "{\em cycle
extraction}" and "{\em link recovery}" have to be used. The
process of cycle extraction and link recovery are described as
follows.

\begin{figure}[htbp]
 \center
 \scalebox{0.5}{\includegraphics{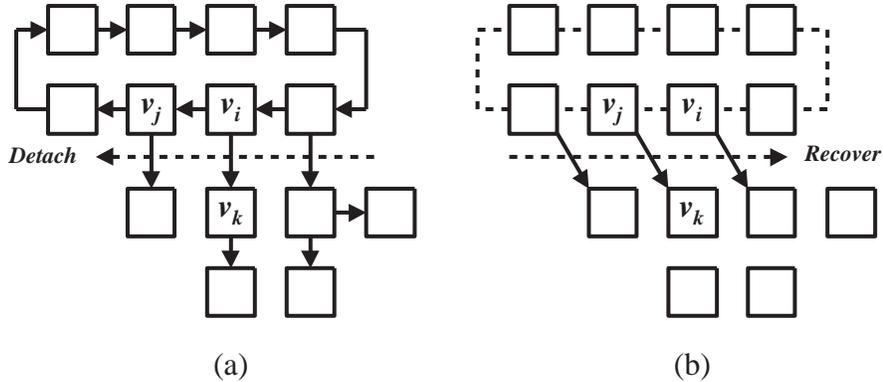}}
 \caption{The process of (a) cycle extraction and (b) link recovery.}
 \label{figure13}
\end{figure}

{\bf Cycle Extraction}: A forest basically contains one and only
one cycle $G^C=(V^C,E^C)$ as a sub-digraph with a subset of $V^C$
linked to either a null node, the head of a queue, or the root of
a tree. The process of cycle extraction detaches all the null
nodes, queues, and trees from the cycle by cutting all the edges
in $E=\{\overrightarrow{v_i v_j} \mid v_i \in V^C, v_j \notin
V^C\}$, as shown in Fig.\ref{figure13}(a). This will transform a
forest into one cycle (arrow T1 in Fig.\ref{figure12}) and a
collection of null nodes, queues, and/or trees (arrow T2, T3, and
T4 respectively). Each of the null nodes and queues can further be
transformed into loopbacks and cycles via process E1 and E2. If
there are still any trees in the remaining digraph, extensions can
be made again to transform the trees into forests (process E3) and
the procedure of cycle extraction can be applied recursively
(arrow T1, T2, T3, and T4) until no trees are left. This procedure
eventually transforms a forest into loopbacks and/or cycles, so
that the permutation can be implemented using $6$ layers of {\bf
\small CN} gates in parallel.

{\bf Link Recovery}: After each cycle has been implemented, the
links that had been cut must be recovered. That is, for each
$\overrightarrow{v_i v_j} \in E^C$, if $\overrightarrow{v_i v_k}
\in E$ but $\overrightarrow{v_i v_k} \notin E^C$, $v_j$ must be
replicated to $v_k$, as shown in Fig.\ref{figure13}(b). Since
there will be at most $n-2$ such $k$'s in a multicast connection
digraph, in the worst case the replication can be done by
$\log_2n$ layers of {\bf \small CN} gate. This completes the
implementation of a forest. For a tree, it can be extended to a
forest via process E3 and then follow the algorithm to do further
reduction in the same way.

\subsubsection{Unicast Connection Digraph}

\begin{figure}[htbp]
 \center
 \scalebox{0.4}{\includegraphics{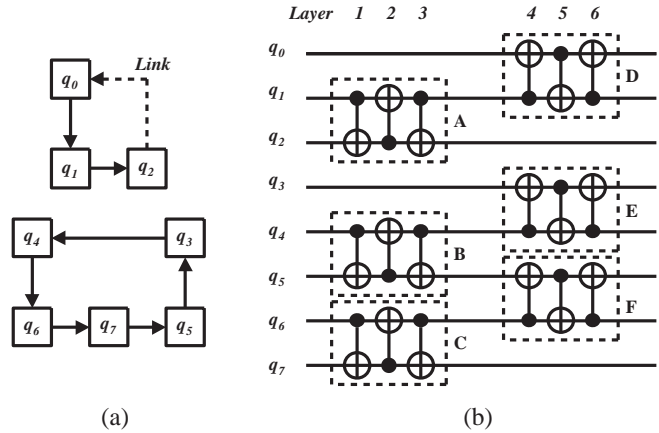}}
 \caption{(a) A unicast connection digraph, and (b) its quantum circuits}
 \label{figure14}
\end{figure}

Following the guideline described above, in this section we show
how a unicast connection digraph can be implemented with a time
complexity of $O(1)$ and a space complexity of $O(n)$. A typical
unicast connection status at a given time is shown by the solid
arrows in Fig.\ref{figure14}(a). The switching module needs to
perform two connection sub-digraphs:
\begin{eqnarray}
G^C & = & (q_3,q_4,q_6,q_7,q_5),\\
G^Q & = & [q_0,q_1,q_2].
\end{eqnarray}
These can be done by first extending $G^Q$ to
$G^{C^\prime}=(q_0,q_1,q_2)$, as shown by the dash link in
Fig.\ref{figure14}(a), and then implement $G^C$ and $G^{C^\prime}$
using $6$ layers of {\bf \small CN} gates. As described
previously, the sub-digraph $G^C=(q_3,q_4,q_6,q_7,q_5)$ can be
done by first applying
\begin{equation}
X=(q_6,q_7)(q_4,q_5)
\end{equation}
and then
\begin{equation}
Y=(q_6,q_5)(q_4,q_3).
\end{equation}
The transposition $(q_4,q_5)$ is done by
\begin{equation}
(q_4,q_5) = CN(q_4,q_5) \cdot CN(q_5,q_4) \cdot CN(q_4,q_5),
\end{equation}
as shown by block B in Fig.\ref{figure14}(b). In the same way,
$(q_6,q_7)$, $(q_4,q_3)$, $(q_6,q_5)$ are done by blocks C, E, F
respectively. Similarly, the implementation of
$G^{C^\prime}=(q_0,q_1,q_2)$ can be done by first applying
$X=(q_1,q_2)$ and then $Y=(q_1,q_0)$. These are implemented as
blocks A and D in Fig.\ref{figure14}(b).

Note that, independent of the switch size $n$, the whole circuit
can be completed in $6$ layers of {\bf \small CN} gates over $n$
qubits. This achieves a time complexity of $O(1)$ and a space
complexity of $O(n)$.

\subsubsection{Multicast Connection Digraph}

\begin{figure}[htbp]
 \center
 \scalebox{0.4}{\includegraphics{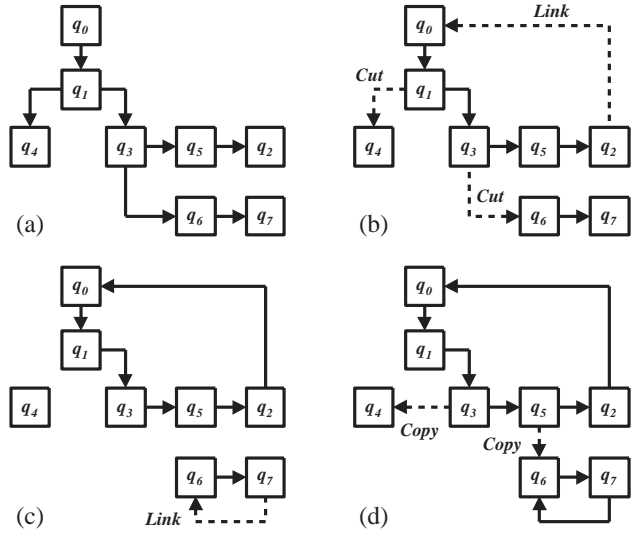}}
 \caption{Procedures for implementing a multicast connection digraph.}
 \label{figure15}
\end{figure}

In classical packet switching, the input packets are usually
buffered in the memory, multicasting can be easily achieved by
reading the packet once and writing the same packet to multiple
destinations. If the switching is done in the quantum domain,
multicasting can be done by replicating the input qubit to
multiple destination qubits. A typical multicasting configuration
is shown in Fig.\ref{figure15}(a). In this example, the switching
module needs to perform the following connection digraph:
\begin{equation}
G^T=[q_0,q_1][q_1,q_4][q_1,q_3][q_3,q_5,q_2][q_3,q_6,q_7].
\end{equation}

\begin{figure}[htbp]
 \center
 \scalebox{0.4}{\includegraphics{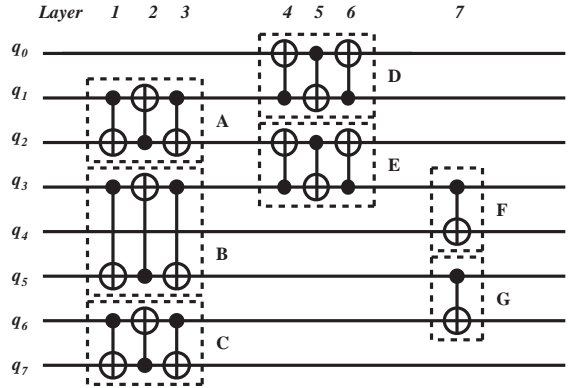}}
 \caption{Quantum circuits for a multicast connection digraph.}
 \label{figure16}
\end{figure}

Following the guideline, each of the steps is shown below:
\begin{enumerate}
\item  The tree $G^T$ can be extended to a forest by linking
any leaf, say $q_2$, to $q_0$. The cycle extraction procedure is
then performed to cut $\overrightarrow{q_1 q_4}$ and
$\overrightarrow{q_3 q_6}$ down. The result is shown in
Fig.\ref{figure15}(b).
\item  The extension and cycle extraction
processes are recursively applied to $[q_6,q_7]$ until no tree is
left, as shown in Fig.\ref{figure15}(c).
\item Each of the disjointed sub-digraphs can be implemented
in parallel. The sub-digraph $G^C=(q_0,q_1,q_3,q_5,q_2)$ can be
done by first applying $X=(q_1,q_2)(q_3,q_5)$ and then
$Y=(q_1,q_0)(q_3,q_2)$, while $G^{C^\prime}=(q_6,q_7)$ can be
implemented directly, as shown by blocks A, B, D, E, and C in
Fig.\ref{figure16}.
\item Each of the disconnected edges has to be recovered,
so $q_3$ needs to be replicated to $q_4$, and $q_5$ needs to be
replicated to $q_6$, as shown in Fig.\ref{figure15}(d). These can
be done by blocks F and G in Fig.\ref{figure16}.
\end{enumerate}

In general, the total number of layers for implementing a
multicast connection digraph is $6+\lceil\log_2(r+1)\rceil$, where
$r$ is the maximum number of $\overrightarrow{v_j v_k}$
($k=1,2,\ldots r$) that are to be recovered. In the worst case,
when one inlet is broadcast to all other $n-1$ outlets, the whole
connection digraph can be done in $O(\log_2n)$ layers of {\bf
\small CN} gates over $n$ qubits. This results in a time
complexity of $O(\log_2n)$ and a space complexity of $O(n)$.

\subsection{Advantages of Quantum Switching}

The advantages of performing digital switching in the quantum
domain are summarized as follows. First, switching in the quantum
domain is {\em strict-sense non-blocking}. A switch is called
strict-sense non-blocking if the network can always connect each
idle inlet to an arbitrary idle outlet independent of the current
network permutation \cite{Pat98}. Note that switching in the space
domain is not always non-blocking. Sometimes, the required data
path can not be established even if the output port is available.
It has been shown that for an $n \times n$ network in
Fig.\ref{figure5}(a)) to be non-blocking, there must be at least
$2n-1$ modules in the middle stage \cite{Clo53}. However,
switching in the quantum domain is actually a unitary
transformation, which is always possible. This results in the fact
that a quantum switch is non-blocking in the strict sense.

Second, it takes only $n$ qubits to build a quantum switch, the
space complexity is $O(n)$ in terms of the number of qubits. The
problem of space complexity is an important issue in the classical
space switching. To make a classical space switch non-blocking, a
certain number of modules in the middle stage have to be used to
allocate a physical path for each connection, so the number of
cross-points increases with the size of the switch. For example,
with optimal grouping, the minimum number of cross-points for the
switch shown in Fig.\ref{figure5}(a) is $N_{min}=4n(\sqrt{2n}-1)$,
where $n$ is the total number of inlets/outlets \cite{Clo53}.
However, an $n \times n$ quantum switch uses only $n$ qubits as
the basis to perform the switching, which is a reasonable resource
consumption.

Third, quantum switching is scalable in terms of time complexity.
In a classical time switch, usually the bottleneck is the speed of
the switching device. Because when the throughput increases, the
time duration for switching a particular bit of data decreases.
For example, in a memory switch with throughput $T$, the memory
speed must be at least $1/2T$ to allow one read and one write
operation to be performed. However, in the quantum switching, the
time complexity is not sensitive to the throughput. A high
throughput quantum switch can be achieved simply by increasing the
number of I/O ports, which only induces a reasonable amount
($O(n)$) of space consumption. However, even in the worst case
scenario, the throughput gain still outweights the time penalty in
a classical time domain switch ($O(n)$ v.s. $O(\log_2n)$).

\section{Conclusions}

Networks are rapidly growing due to increased number of users and
rising demands for bandwidth-intensive services. To support such a
huge traffic volume, a wide range of different technologies are
being proposed as the core of a high performance switch. In this
paper, an architecture of digital quantum switching is presented.
The proposed mechanism allows digital data to be switched using a
series of quantum operations. The procedures of how to implement
unicast and multicast connections are discussed in detail. In
terms of the blocking rate, this architecture is strict-sense
non-blocking. From a complexity point of view, the space
complexity grows only linearly with the number of I/O ports, and
the time complexity is constant for unicasting and logarithmic for
multicasting. This architecture is suitable for deploying high
throughput switching devices so that high bandwidth demand can be
met.

\end{document}